\documentclass[%
reprint,
superscriptaddress,
 amsmath,amssymb,
pra,
]{revtex4-2}

\usepackage{graphicx}
\usepackage{dcolumn}
\usepackage{bm}
\usepackage{amsmath}
\usepackage{amssymb}
\usepackage[utf8]{inputenc}
\usepackage[T1]{fontenc}
\usepackage{mathptmx}
\usepackage{xcolor}
\usepackage{enumitem}   
\usepackage{subfig}
\usepackage[format=plain,labelfont=up,textfont=up]{caption}
\usepackage{dsfont}
\usepackage{xr-hyper}
\usepackage{hyperref}
\usepackage{url}
\usepackage{siunitx}
\usepackage{float}
\usepackage{soul,xcolor}

\usepackage{textgreek}
\usepackage{upgreek}

\begin{document}

\title{Polarization-dependent effects in vibrational absorption spectra of 2D finite-size adsorbate islands on dielectric substrates}

\author{Benedikt Zerulla*}
\affiliation{Institute of Nanotechnology,
Karlsruhe Institute of Technology (KIT),
Kaiserstr. 12, 76131 Karlsruhe, Germany}
\email{benedikt.zerulla@kit.edu\\christof.woell@kit.edu\\carsten.rockstuhl@kit.edu}
\author{Marjan Krstić}
\affiliation{Institute of Theoretical Solid State Physics,
Karlsruhe Institute of Technology (KIT),
Kaiserstr. 12, 76131 Karlsruhe, Germany}
\author{Shuang Chen}
\affiliation{Institute of Functional Interfaces,
Karlsruhe Institute of Technology (KIT),
Kaiserstr. 12, 76131 Karlsruhe, Germany}
\author{Zairan Yu}
\affiliation{Institute of Functional Interfaces,
Karlsruhe Institute of Technology (KIT),
Kaiserstr. 12, 76131 Karlsruhe, Germany}
\author{Dominik Beutel}
\affiliation{Institute of Theoretical Solid State Physics,
Karlsruhe Institute of Technology (KIT),
Kaiserstr. 12, 76131 Karlsruhe, Germany}
\author{Christof Holzer}
\affiliation{Institute of Theoretical Solid State Physics,
Karlsruhe Institute of Technology (KIT),
Kaiserstr. 12, 76131 Karlsruhe, Germany}
\author{Markus Nyman}
\affiliation{Institute of Nanotechnology,
Karlsruhe Institute of Technology (KIT),
Kaiserstr. 12, 76131 Karlsruhe, Germany}
\author{Alexei Nefedov}
\affiliation{Institute of Functional Interfaces,
Karlsruhe Institute of Technology (KIT),
Kaiserstr. 12, 76131 Karlsruhe, Germany}
\author{Yuemin Wang}
\affiliation{Institute of Functional Interfaces,
Karlsruhe Institute of Technology (KIT),
Kaiserstr. 12, 76131 Karlsruhe, Germany}
\author{Thomas G. Mayerhöfer}
\affiliation{Leibniz Institute of Photonic Technology,
Albert-Einstein-Str. 9, 07745 Jena, Germany}
\author{Christof Wöll*}
\affiliation{Institute of Functional Interfaces,
Karlsruhe Institute of Technology (KIT),
Kaiserstr. 12, 76131 Karlsruhe, Germany}
\email{christof.woell@kit.edu}
\author{Carsten Rockstuhl*}
\affiliation{Institute of Nanotechnology,
Karlsruhe Institute of Technology (KIT),
Kaiserstr. 12, 76131 Karlsruhe, Germany}
\affiliation{Institute of Theoretical Solid State Physics,
Karlsruhe Institute of Technology (KIT),
Kaiserstr. 12, 76131 Karlsruhe, Germany}
\email{carsten.rockstuhl@kit.edu}

\begin{abstract}
In the last years, Infrared Reflection-Absorption Spectroscopy (IRRAS) became a standard technique to study vibrational excitations of molecules. These investigations are strongly motivated by perspective applications in monitoring chemical processes. For a better understanding of the adsorption mechanism of molecules on dielectrics, the polarization-dependence of an interaction of infrared light with adsorbates at dielectric surfaces is commonly used. Thus, the peak positions in absorption spectra could be different for s- and p-polarized light. This shift between the peak positions depends on both the molecule itself and the dielectric substrate. While the origin of this shift is well understood for infinite two-dimensional adsorbate layers, finite-size samples, which consist of 2D islands of a small number of molecules, have never been considered. Here, we present a study on polarization-dependent finite-size effects in the optical response of such islands on dielectric substrates. The study uses a multi-scale modeling approach that connects quantum chemistry calculations to Maxwell scattering simulations. We distinguish the optical response of a single molecule, a finite number of molecules, and a two-dimensional adsorbate layer. We analyze CO and CO$_2$ molecules deposited on CeO$_2$ and Al$_2$O$_3$ substrates. The evolution of the shift between the polarization-dependent absorbance peaks is firstly studied for a single molecule, which it does not exhibit for at all, and for finite molecular islands, which it increases with increasing island size for, as well as for an infinite two-dimensional adsorbate layer. In the latter case, the agreement between the obtained results and the experimental IRRAS data and more traditional three/four-layer-model theoretical studies supports the predictive power of the multi-scale approach.
\end{abstract}

\maketitle

\section{Introduction and summary}
 \begin{figure*}[t!]
\centering
     
	\includegraphics[width=0.4\textwidth]{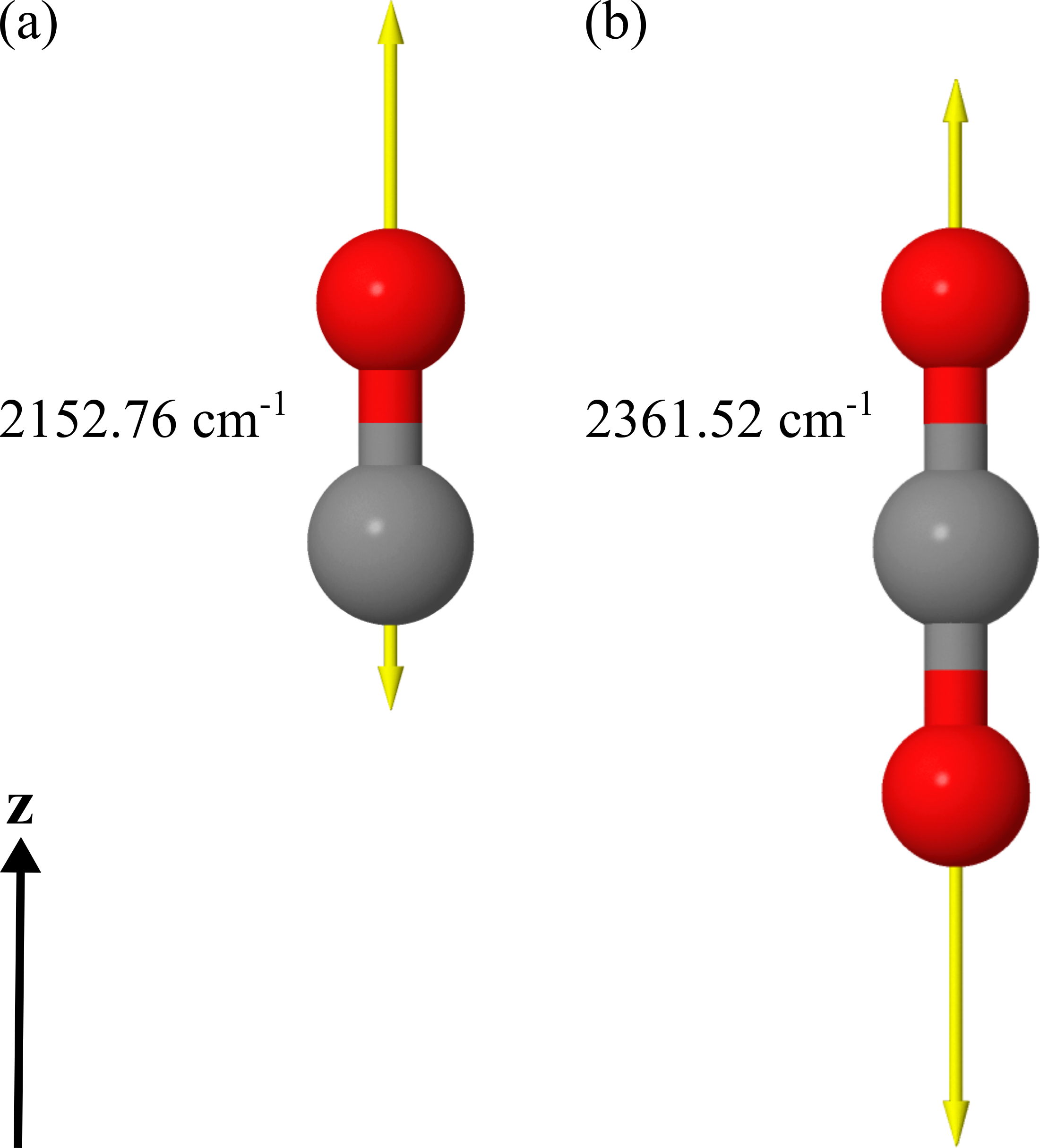}
\hspace{0cm}
	\caption{\textbf{(a)} DFT optimized CO molecule with the displacement vector of the stretching vibrational mode at 2152.76 cm\textsuperscript{-1}. The calculated vibrational wavenumber was scaled by 0.962 to match the experimental value. The molecule is oriented along the z-axis in the calculation. \textbf{(b)} DFT optimized CO\textsubscript{2} molecule with the displacement vector of the anti-symmetric stretch vibrational mode at 2361.52 cm\textsuperscript{-1}. The calculated vibrational wavenumber was scaled by 0.962 to match the experimental value. The molecule is oriented along the z-axis in the calculation. }
    \label{fig:CO_CO2}
	\end{figure*}
Infrared Reflection-Absorption Spectroscopy (IRRAS) is one of the most important experimental techniques for characterizing molecular species on solid surfaces. \cite{GOLDEN1981395, Greenler1966, HOFFMANN1983107, Molt1984} Its application to metal surfaces is a rather mature field, with extensive experimental data for various molecules for virtually all metals in the periodic table. \cite{Greenler1966, Trenary2000, Hinrichs2011, Zaera2014} There is also a well-advanced theoretical understanding through state-of-the-art \textit{ab initio} calculations concerning binding energies and vibrational frequencies. \cite{DEGLMANN2002511, Deglmann20029535}

In contrast, the application of IRRAS to oxides and dielectrics, in general, is much less developed. While experimental IR data for numerous adsorbates on oxide powder samples are widely available, \cite{Wang2017, doi:10.1080/23746149.2017.1296372} IRRAS data for macroscopic single crystals of these materials are limited. \cite{Wang2017,doi:10.1080/23746149.2017.1296372,HEAD201751, Zhang2022} This limitation originates from the particular optical properties of dielectrics and has hindered the experimental recording of IRRAS data on oxides until recently. The key difference between metals and semiconductors is the screening of the electric field by metal electrons, affecting the total infrared reflectivity and giving rise to the so-called surface selection rule governing IRRAS on metal surfaces. \cite{Greenler1966, GREENLER1982415} This rule states that for metals, usually, only vibrations with a component of the transition dipole moment (TDM) normal to the surface are detectable in IRRAS, while those parallel are screened by the metal electrons. This restriction does not apply to oxides and dielectrics in general, allowing IRRAS data collection using both s- and p-polarized light. \cite{CODipExp} Notably, IRRAS data for dielectrics show unique, rather counterintuitive characteristics, such as a splitting of peaks recorded for different polarizations. \cite{CODipExp,doi:10.1021/acs.jpcc.1c10181}

For weakly absorbing adsorbates, the variation in absorbance peak positions between s- and p-polarized light is minimal, often just a few wavenumbers. Reports on these polarization-dependent shifts are limited due to experimental challenges. Notably, Chabal and colleagues observed minor shifts in Si-H on Si(100) substrates, \cite{PhysRevLett.54.1055} while the first data for oxides were reported by Yang, {\it et al.} \cite{CODipExp} demonstrating significant effects for CO on ceria and another dielectric material, CaCO$_3$. Moreover, for strong absorbers such as N$_2$O, this shift can reach up to $20\,\mathrm{cm}^{-1}$. \cite{doi:10.1021/acs.jpcc.1c10181}    

It is possible to study chemical processes by means of a combination of \textit{in situ} spectroscopic methods and theoretical simulations. An instance is the investigation of metal nanoparticle oxidation with plasmonic optical spectroscopy in Refs. \cite{doi:10.1021/acs.jpcc.9b00323,C9NR07681F}. Recent developments in infrared spectroscopy have demonstrated the potential of IR spectroscopy, in particular Diffuse Reflectance IR Spectroscopy (DRIFTS) for monitoring chemical processes on oxide powder surfaces \cite{doi:10.1021/acscatal.9b04016} especially for operando studies \cite{doi:10.1021/acs.jpcc.3c03567} in the context of heterogeneous catalysis. A crucial aspect is determining whether polarization-dependent shifts observed for adsorbate vibrational frequencies on single crystals are also present on smaller oxide particles, which could impact the interpretation of powder data. Traditional theoretical models, such as the three-layer-model, \cite{CODipExp} have successfully replicated experimental findings on macroscopic single crystal surfaces but are inadequate for studying finite-size effects, as they assume a dielectric permittivity for a molecular layer being infinitely extended into two dimensions. For a molecular island which consists of only a small number of molecules, this assumption does not hold.

To address this lack of understanding of finite-size effects, we have adopted here a multi-scale \textit{ab initio} modeling approach which connects density-functional theory (DFT) calculations to self-consistent scattering simulations based on Maxwell's equations, enabling the examination of an extended 2D adsorbate layer, or adlayer in brief, and finite-sized adsorbate islands. This approach aims to better understand the polarization-dependent shift and its relevance to different-sized particles, enhancing the interpretation of experimental data in real-world applications. 

In Ref.~\cite{CODipExp}, it is explained on a macroscopic level that the reflectance of s-polarized light is determined by the imaginary part of the permittivity of the molecular film associated with the molecular axis. In contrast, the reflectance of the electric field component of the p-polarized light, which is vertical to the film, is determined by the imaginary part of the inverse of this permittivity. The difference between the maxima of both quantities explains the shift between the curves. The electric field component of the p-polarized light, which is vertical to the film, increases with the angle of incidence, so that the reflectance of p-polarized light is mainly determined by this component.

Chabal presented that a dielectric permittivity of a 2D adsorbate layer can be phenomenologically modeled. \cite{CHABAL1988211} This can be further used to compute the optical response of the layer deposited on a substrate. While infinite systems can be described by phenomenological means, as reported in Ref.~\cite{CODipExp}, one cannot use material parameters such as the dielectric permittivity for finite islands consisting only of a few molecules.

Here, we demonstrate with our multi-scale \textit{ab initio} modeling approach for CO and CO$_2$ molecules and two different dielectric substrates that the shifts between the curves for s- and p-polarized light, related to the different reflection conditions for the s-polarized light and the vertical electric field component of the p-polarized light, increase with an increasing size of the finite islands. In this regard, it is possible to explain the origin of the shift and to observe the transition from the shift for a single molecule and for small islands toward the shift for the case of the infinite adsorbate layer.

The article is structured as follows. In Section~2,  the DFT simulations of the molecules are discussed. In Section~3, the optical response of islands of CO and CO$_2$ molecules on dielectric substrates is analyzed.  The conclusions of the article are presented in Section~4. Technical details are presented in the following three sections that serve as appendix. In Appendix~A, the experimental measurements are described. Appendix~B explains the T-matrix formalism, and Appendix~C describes the three/four-layer-model study.
\section{DFT simulations of CO and CO\textsubscript{2} molecules}
We apply a dedicated multi-scale approach to study the optical response of molecular islands on dielectric substrates. To this end, we calculate the optical properties of individual molecules on the quantum level using time-dependent density functional theory. From these considerations, we assign a polarizability to the molecules. By representing the individual molecule using the polarizability, we use a full-wave Maxwell solver based on a multiple-particle scattering formalism to study the optical response of an infinite periodic arrangement of molecules on a substrate or the response of a finite island of molecules. The finite 2D islands are also periodically arranged into a superlattice, but only for numerical convenience. The lattice constant of the superlattice will be sufficiently large to suppress interaction among the periodically arranged islands and sufficiently small so that diffraction effects do not play a role. Comparing the simulations to experiments, on the one hand, permits us to identify the microscopic origin of the different responses between s- and p-polarized light at oblique incidence studied in the past. \cite{PhysRevLett.54.1055,CODipExp,doi:10.1021/acs.jpcc.3c03567} On the other hand, by comparing the response between the infinite adsorbate layer and the finite islands, we can pinpoint the spatial extent of an adsorbate island necessary to consider it sufficiently representative of the bulk sample.    

In our examples, we use DFT to initially optimize the geometry of the molecular structure in the minimum on the potential energy surface of the electronic ground state (presented in Figures~\ref{fig:CO_CO2}\textbf{(a)} and \textbf{(b)}). Once the structure is obtained, we calculate the vibrational and electronic dynamic polarizabilities of the CO and CO\textsubscript{2} molecules in the gas phase using a linear-response approach within the time-dependent density functional theory (TDDFT) framework. In the considered spectral domain (mainly 2100-2200 cm\textsuperscript{-1} for CO and 2300-2500 cm\textsuperscript{-1} for CO\textsubscript{2}), the vibrational dynamic polarizability is characterized by a single resonance in the z-component of the polarizability in the chosen coordinate system. The displacement vectors of that vibrational mode are also visualized in Figure~\ref{fig:CO_CO2} for each molecule  along with an indication of its resonance frequency.

\begin{figure*}[t!]
\centering
     \subfloat{
	\includegraphics[width=0.45\textwidth]{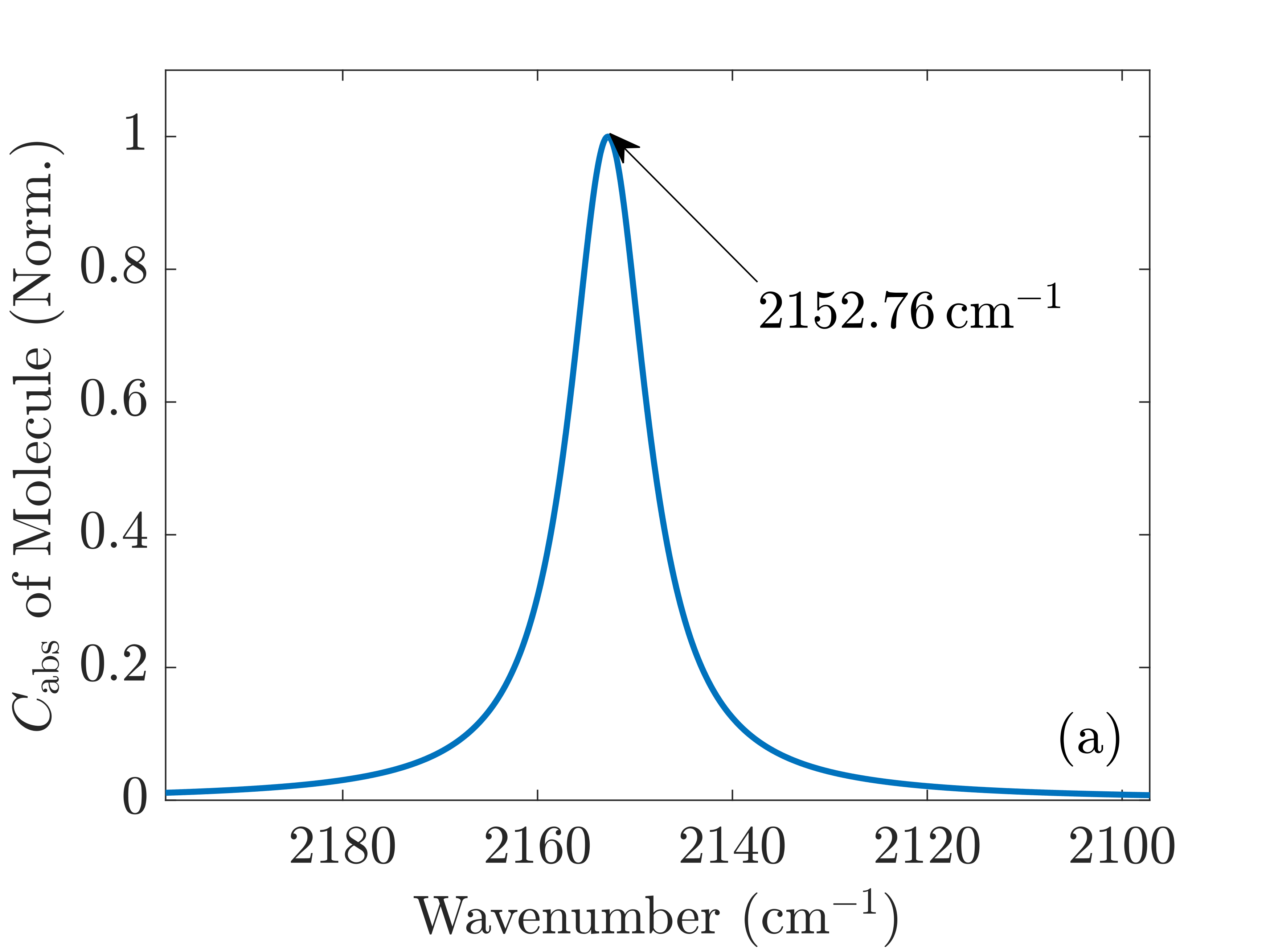}
 }\hspace{0cm}
 \subfloat{
	 \includegraphics[width=0.45\textwidth]{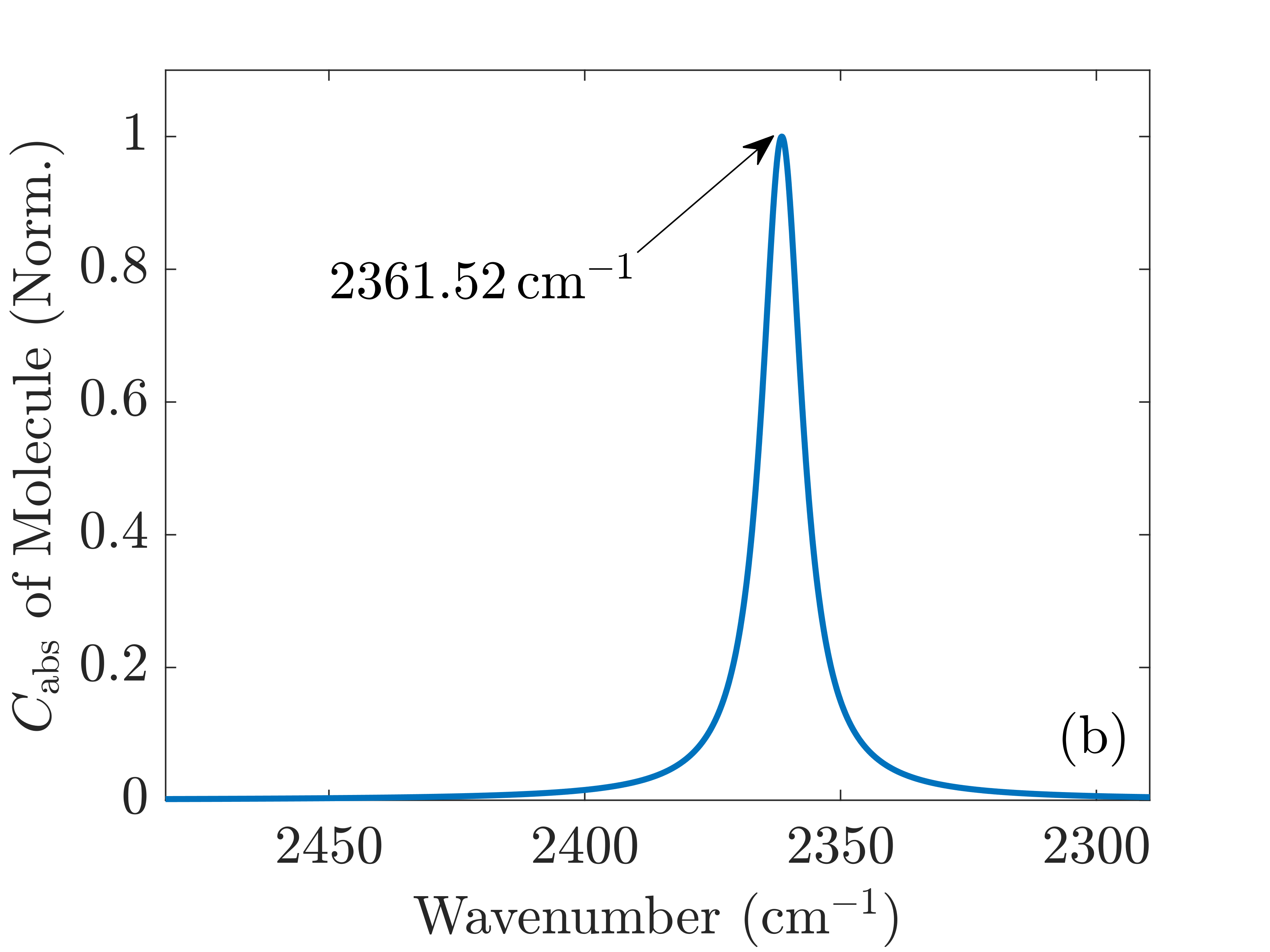}
}
	\caption{\textbf{(a)} Normalized absorption cross sections of a single CO and \textbf{(b)} of a single CO$_2$ molecule in the gas phase with respect to a plane wave with a z-polarized electric field. Both molecules are aligned along the z-axis, so their associated absorption is resonantly enhanced only for a z-polarized incident field.}
    \label{fig:AbsMolecule}
	\end{figure*}

In the DFT calculations, performed with the development version of TURBOMOLE, \cite{Franzke.Holzer.ea:TURBOMOLE.2023} we combined the hybrid PBE0 exchange-correlation functional \cite{adamoReliableDensityFunctional1999, ernzerhofAssessmentPerdewBurke1999} and the def2-TZVP basis set. \cite{weigendBalancedBasisSets2005} The complete damped dynamic polarizability tensors are calculated around the wavenumbers of the stretching vibrational mode for each molecule. In the case of CO, we calculated all polarizability tensors using the DFT/TDDFT method for the range 2180-2285 cm\textsuperscript{-1} with a resolution of 0.1 cm\textsuperscript{-1}. Calculations were done for all frequencies in the full spectral range from 1 to 4000 cm\textsuperscript{-1} for the CO\textsubscript{2} molecule with the same resolution within the same methods. The damping was set to 5 cm\textsuperscript{-1} for the half-width at half-maximum of the Lorentzian line shape to simulate the broadening seen in the experiments. Due to the known fact that DFT simulations of single molecules in the gas phase overestimate the frequencies of the vibrational modes of the molecules, all wavenumbers of the spectra were scaled down by a factor of 0.962 to match the experimental values. \cite{scalefactor} The scaling shifted the spectral window for the CO molecule to the range covering 2097.16-2198.17 cm\textsuperscript{-1} as visualized in Figure~\ref{fig:AbsMolecule}\textbf{(a)}. For the CO\textsubscript{2} molecule, we focused only on the spectral part presented in Figure~\ref{fig:AbsMolecule}\textbf{(b)} after the scaling of wavenumbers was applied. The infrared absorption and the electronic background calculated by DFT in the spectral domain of interest are presented in Figures~S1\textbf{(a)} and \textbf{(b)} in the Supplementary Information. 

Finally, the obtained dynamic polarizability tensors are used to construct the T-matrices of an individual molecule. The T-matrix serves as the complete representation of the optical properties of a molecule. This T-matrix can then be used in a full-wave Maxwell solver to simulate the optical response of an infinite or periodically arranged finite island of molecules on a substrate. Please note that only the optical interaction between the molecules themselves and between the molecules and the substrate is considered in these simulations. These multi-scale simulations rely on a previously published workflow. \cite{SURMOFCavity,Fernandez-Corbaton:2020,https://doi.org/10.1002/adfm.202301093} The simulation of the optical response of the system is performed with the T-matrix-based program suite treams. \cite{Beutel:21,treams,BEUTEL2024109076}

Once the computational workflow is clear, we study the optical response of individual molecules in the gas phase and, afterward, the optical response from infinite or finite islands on a substrate. The configurations, we consider, strictly correspond to those considered in experiments already presented in the literature. \cite{CODipExp}

\section{Islands of CO and CO$_2$ on dielectric substrates}
We apply in this article, for the first time, a multi-scale optical modeling approach \cite{SURMOFCavity,Fernandez-Corbaton:2020,https://doi.org/10.1002/adfm.202301093} to study finite-size effects on the optical response of molecular islands supported by a dielectric substrate. The starting point for describing the optical response of the molecules is their T-matrix.

\subsection{Isolated molecules in the gas phase}

The T-matrices of isolated molecules in the gas phase calculated with DFT express the vibrational $\mathbf{T}_{\mathrm{vib}}$ and the electric $\mathbf{T}_{\mathrm{el}}$ response of the respective molecule in infrared wavelength regions. Although the molecules do not show electronic transitions in this spectral range, the less-dispersive contribution of transitions from the visible to UV range has to be considered via 
\begin{align}
\mathbf{T}=\mathbf{T}_{\mathrm{vib}}+\mathbf{T}_{\mathrm{el}}\,.    
\end{align}
It should be noted that translational and rotational molecular movements are excluded from the molecular degrees of freedom in the theoretical description. In Figures~\ref{fig:AbsMolecule}\textbf{(a)} and \textbf{(b)}, the absorption cross sections of a single CO and CO$_2$ molecule with respect to a plane wave with a z-polarized electric field are shown, respectively. They are calculated from the extinction cross section $C_{\mathrm{ext}}$ and the scattering cross section $C_{\mathrm{sca}}$,
\begin{align} 
C_{\mathrm{abs}}=C_{\mathrm{ext}}-C_{\mathrm{sca}}\, ,
\end{align}
with 
\begin{align}
    C_{\mathrm{ext}}=-\frac{1}{k^2}\sum_{m=-1}^1\mathrm{Re}\left(a_{1m,\mathrm{N}}c^*_{1m,\mathrm{N}}+a_{1m,\mathrm{M}}c^*_{1m,\mathrm{M}}\right)\, ,
\end{align}
see Equation~(5.18b) from Ref.~ \cite{MishBook}, and 
\begin{align}
    C_{\mathrm{sca}}=\frac{1}{k^2}\sum_{m=-1}^1\left(|c_{1m,\mathrm{N}}|^2+|c_{1m,\mathrm{M}}|^2\right)\, ,
\end{align}
see Equation~(5.18a) from Ref.~ \cite{MishBook}. $a_{1m,\mathrm{N}}$ and $a_{1m,\mathrm{M}}$ are the multipolar expansion coefficients of the incident plane wave, while $c_{1m,\mathrm{N}}$ and $c_{1m,\mathrm{M}}$ are the multipolar expansion coefficients of the scattered wave, see Equation~(\ref{eq:EExpan}). 
Moreover, we show the absorption only in a relevant spectral domain close to their respective absorption wavenumbers. We observe that a clear resonant response emerges. Note that when excited with an electric field along either the x- or the y-direction, the non-dispersive background causes a small but finite response. These are the low-energy shoulders of electronic transitions occurring in the ultraviolet part of the spectrum. Even though these contributions play the role of an almost constant background in the spectral range where the vibrations are present, they need to be considered to compute the intensity of the spectra correctly. 

\subsection{Islands of CO molecules supported on a substrate}

Once we know the spectral properties of the individual molecules, we can study the optical response after they are deposited as islands on top of a substrate. In this subsection, we discuss the case of CO, and in the following subsection, we discuss CO$_2$.

\begin{figure*}[t]
\centering
     
	\includegraphics[width=0.8\textwidth]{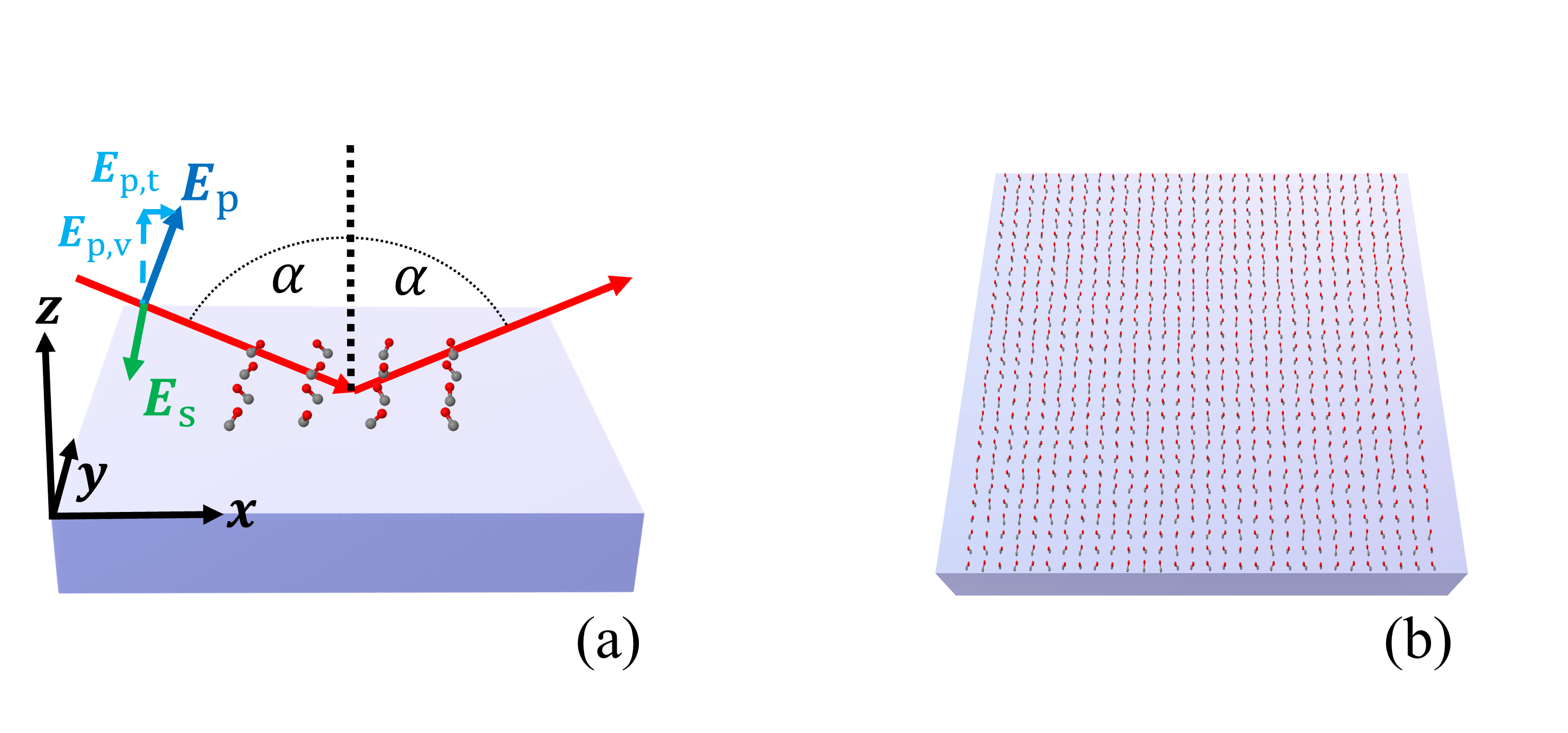}

	\caption{\textbf{(a)} Setup of simulation and measurement. A plane wave, either s- or p-polarized, illuminates the finite molecular adsorbate islands on top of the dielectric substrate under oblique incidence with an angle $\alpha$. \textbf{(b)} Schematic representation of the infinite 2D adsorbate layer.}
    \label{fig:islandImages}
	\end{figure*}

The numerical setup shown in Figure~\ref{fig:islandImages}\textbf{(a)} is constructed in analogy to the experimental setup to record IRRAS data for adsorbates on oxide single crystal surfaces. \cite{Wang2017,doi:10.1080/23746149.2017.1296372} The substrate with the molecular adsorbate islands is illuminated from above with a plane wave under oblique incidence in the xz-plane. The angle of incidence was set to $\alpha$ = 80$^{\circ}$. The incident plane wave is either s- or p-polarized. In our coordinate system, for normal incidence, s- and p- correspond to a polarization of the electric field along the y- and the x-axis, respectively. For oblique incidence, the polarizations are correspondingly rotated.
It has to be noted that in the case of s-polarized light, the electric field of the incident light is perpendicular to the plane of the surface normal and the vector of incidence. In the case of p-polarized light, the electric field consists of a component $\bm{E}_{\mathrm{p,v}}$, which is perpendicular to the surface, and a component $\bm{E}_{\mathrm{p,t}}$, which is parallel to the surface, see Figure~\ref{fig:islandImages}\textbf{(a)}.

We consider the adsorption of CO and CO$_2$ on two different dielectric substrates: CeO$_2$(111), characterized by a dielectric permittivity of 5.31, \cite{CODipExp} and $\alpha$-Al$_2$O$_3$(0001), characterized by a dielectric permittivity of 3.075. \cite{PhysRevB.61.8187} These values correspond to refractive indices of 2.30 and 1.75, respectively. We have assumed constant optical properties for both substrates as the frequencies are far off any resonances. For the value of the dielectric permittivity of Al$_2$O$_3$, we have averaged the values of the principal axes from Ref.~ \cite{PhysRevB.61.8187}, since the anisotropy in this frequency range is very small.   

Figure~\ref{fig:islandImages}\textbf{(b)} schematically shows an infinite 2D adsorbate layer. In Ref.~\cite{CODipExp}, the experimental details of CO molecules on top of a ceria substrate were provided, and we consider here the molecular details as discussed there. Specifically, average tilt angles are assumed, namely $46^{\circ}$ for CO, and of $60^{\circ}$ for CO$_2$. There might be two contributions to this tilt angle, a static tilt and a ``dynamic'' tilt caused by frustrated rotations that are thermally occupied. \cite{CODipExp} The estimated tilt angle for CO$_2$ agrees well with the values found in the three/four-layer-model, as we see in Appendix~A.3. The azimuthal orientation of the molecule is assumed to be random, in agreement with experimental observations. \cite{CODipExp} To accommodate this random orientation in our simulations, we average the T-matrix of the molecule across all possible azimuthal angles to express a nominal molecule on the substrate. We will first consider infinite 2D adlayers with a square lattice and a lattice constant of 0.5\,nm before considering finite islands. We set the distance between the center of mass of the CO molecules and the surface of the substrate to 0.3446\,nm and for the CO$_2$ molecules to 0.3581\,nm. These values were computed from the distance between the center of mass and the atom attached to the surface and an estimate for the distance between this atom and the substrate. 

\begin{figure*}[t]
\centering
	\includegraphics[width=0.9\textwidth]{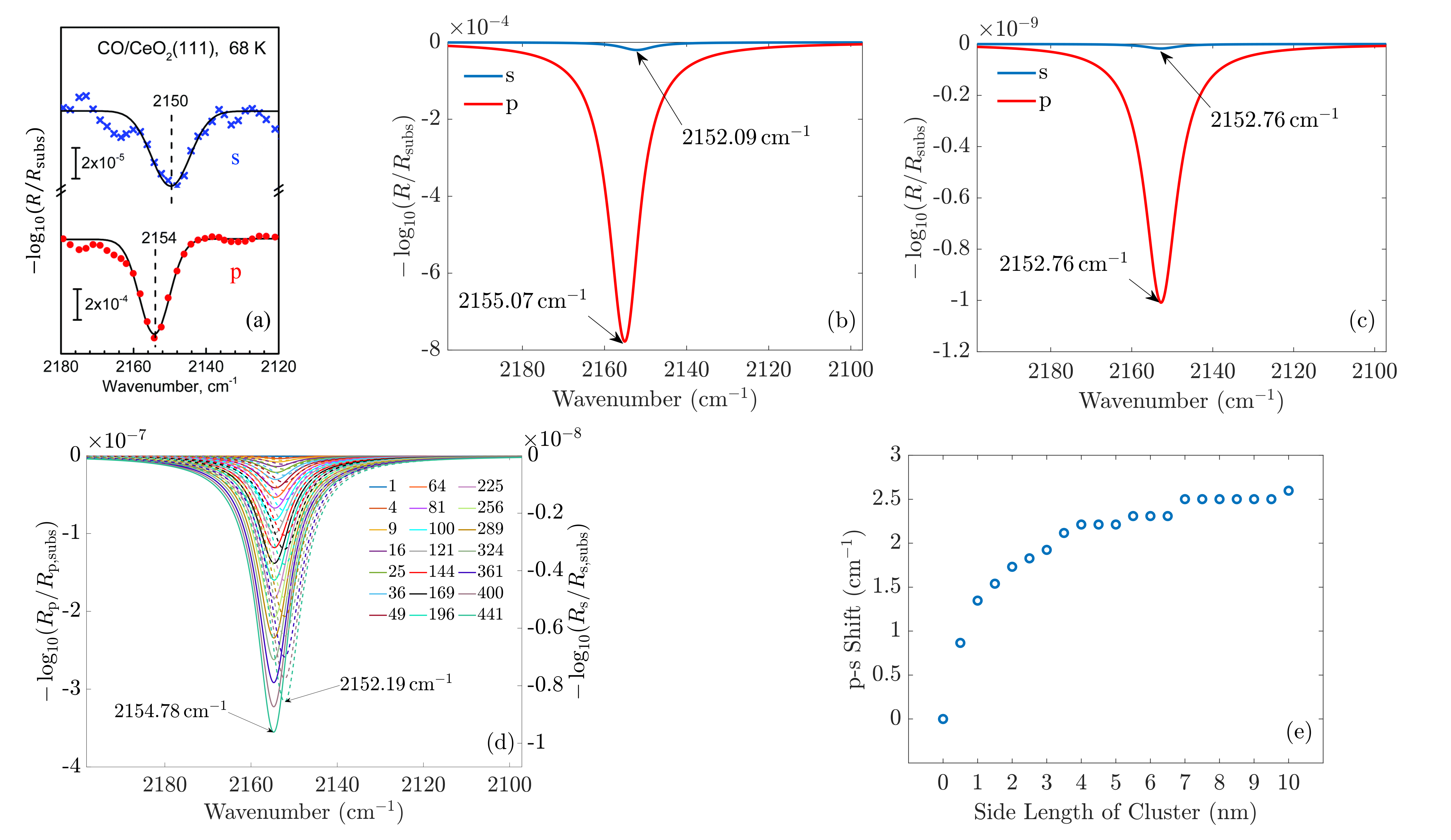}
	\caption{\textbf{(a)} Experimentally measured IRRAS data and \textbf{(b)} simulated spectra of CO molecules on a ceria substrate. The agreement between the measured and the simulated spectra is excellent. \textbf{(c)} and \textbf{(d)} show the simulated spectra of a single molecule on the substrate and for increasing island size, respectively. One observes that the shift between the p- and s-spectra shown in \textbf{(e)} increases from zero for one molecule to 2.6\,cm$^{-1}$. In \textbf{(d)}, the p-spectra are solid lines, and the s-spectra are dashed lines. Fig. 4~ \textbf{(a)} is adapted with permission from \cite{CODipExp} under a \href{https://creativecommons.org/licenses/by-nc/3.0/\#ref-appropriate-credit}{CC BY-NC 3.0 Deed license}. Copyright \textcopyright\ 2020 The Royal Society of Chemistry.}
    \label{fig:CompExpToTheoryCOCeria}
	\end{figure*}

To discuss the spectral details, we define the absorbance $A$ for both polarizations as the difference of the negative logarithms of the reflectance $R$ of the molecules on the substrate and the reflectance $R_{\mathrm{subs}}$ of the bare substrate:
\begin{align}
    A=-\log_{10}(R)+\log_{10}(R_{\mathrm{subs}})=-\log_{10}\left(\frac{R}{R_{\mathrm{subs}}}\right)\,.
\end{align}
To validate our approach, we compare the optical response of CO molecules to previous experimental results and to theoretical work performed using a phenomenological three-layer-model. \cite{CODipExp} In Figures~\ref{fig:CompExpToTheoryCOCeria}\textbf{(a)} and \textbf{(b)}, the experimentally measured IRRAS data and the simulated spectra for CO molecules on the ceria substrate are displayed. Here, we assume initially an infinite periodic array of the molecule. This corresponds to the situation sketched in  Figure~\ref{fig:islandImages}\textbf{(b)}.

In the experiment and the simulation, the optical response is characterized by a resonance in both s- and p-polarization that occurs in spectral proximity to the resonance of the isolated molecules. 

In IRRAS experiments for CO and other molecular species adsorbed on metal surfaces, when the frequency of the IR light matches that of the adsorbed species, the reflectivity always decreases, leading to (positive) peaks in the absorbance spectra. \cite{CHABAL1988211} On dielectric substrates the situation is different. According to the three-layer-model, for s-polarized light, the reflectivity is always increased when hitting the resonance condition, i.e., in normal nomenclature, we would see “negative” absorbance peaks. For p-polarized light, the situation is more complicated, as the electric field consists of a component, which is tangential to the surface, and a component, which is vertical to the surface. The reflectivity can either reduce or increase in resonance, depending on the experimental conditions. \cite{Wang2017}
It is important to note that our multi-scale simulations correctly reproduce the absorbance sign for all the cases studied in this work. In the following, we will always refer to the position and total intensity of these “absorption peaks”.

We see a more intense absorbance peak (in the chosen logarithmic scale) for p-polarization than for s-polarization in both experiment and simulation. Generally, the spectra computed using our new \textit{ab initio} approach are in excellent agreement with the experimental data. As the exact numbers are significant, we have added the information on the spectral position of the resonances in Figure~\ref{fig:CompExpToTheoryCOCeria}.

For CO molecules, the spectral difference between the resonance positions in s- and p-polarization amounts to 3\,cm$^{-1}$. The agreement of the peak positions between the experimentally measured spectra and the results obtained with our multi-scale \textit{ab initio} approach is almost perfect. We note that in contrast to the previous phenomenological description based on the three-layer-model, \cite{CODipExp} free parameters such as the oscillator parameters and the layer thickness were not employed by the present true \textit{ab initio} calculations. Furthermore, in contrast to the macroscopic explanations provided by the three-layer-model, the notion of a bulk permittivity used to describe the molecular film does not enter our consideration. Instead, we consider the optical response at the level of an individual molecule.

In the next step, we consider adsorbate islands of different sizes to study finite size effects. As noted in the introduction, such information is rather important for the interpretation of IR data recorded for CO and CO$_2$ adsorbed on the facets exposed by oxide powder particles with sizes down to the tenths of nm regime. We computed the optical response of finite molecular islands supported on an infinite 2D substrate. Within the variable-sized islands, the molecules continue to be arranged on a quadratic grid with a distance of 0.5\,nm between neighbored molecules in the x- and y-direction, containing between 1 and 441 molecules, with a maximum size of 10×10\,nm$^2$. The T-matrix of the molecules is rotated by the respective tilt angle relative to the z-axis and averaged with respect to the xy-plane. For computational purposes, these square islands on the dielectric substrate are arranged into a periodic two-dimensional (2D) superlattice with a lattice constant of 500\,nm. The large separation from adjacent islands guarantees, first, that the interaction between the islands can be neglected such that each 2D island responds individually and, second, that the s- and p-polarizations and their reflectance continue to be well-defined as the substrate is still infinitely extended. Please note that the chosen period of 500\,nm continues to be much smaller than the wavelengths of interest. Therefore, only a zeroth diffraction order exists for the considered systems.

The simulated spectra for one molecule and for different sizes of the island are shown in Figures~\ref{fig:CompExpToTheoryCOCeria}\textbf{(c)} and \textbf{(d)}, respectively. From these simulations, we extract the spectral position of the absorbance peaks as a function of the island size, see Figure~\ref{fig:CompExpToTheoryCOCeria}\textbf{(d)}.

We observe no shift in peak position between the s- and the p-spectra for a single adsorbate molecule (Figure~\ref{fig:CompExpToTheoryCOCeria}\textbf{(c)}). The absorbance peaks are located precisely at the same spectral position. When there is more than one molecule in the island, a clear shift of the peak position occurs when going from s-polarized light to p-polarized light (Figure~\ref{fig:CompExpToTheoryCOCeria}\textbf{(e)}). This shift increases with growing island size. For 2×2 islands (containing four molecules) with a size of 0.5\,nm × 0.5\,nm, the shift already amounts to $0.9\,\mathrm{cm}^{-1}$. For islands with 441 molecules, with a size of 10\,nm × 10\,nm, the shift is $2.6\,\mathrm{cm}^{-1}$. This value is already relatively close to the results for the infinite adlayer. Increasing the island size, which becomes computationally more and more demanding, would lead to a perfect match between the shift and the positions of the peaks between the finite 2D island and the infinite system. From this study, we conclude that on a microscopic level, the interaction between the molecules is necessary to observe a shift between the s- and p-spectra. For a single molecule, the shift does not occur since this molecule does not interact with another one. Based on these results, one can additionally conclude the number of molecules on average contained in an island by measuring the shift between the s- and p-spectra. 
 \begin{figure*}[t!]
\centering
	\includegraphics[width=\textwidth]{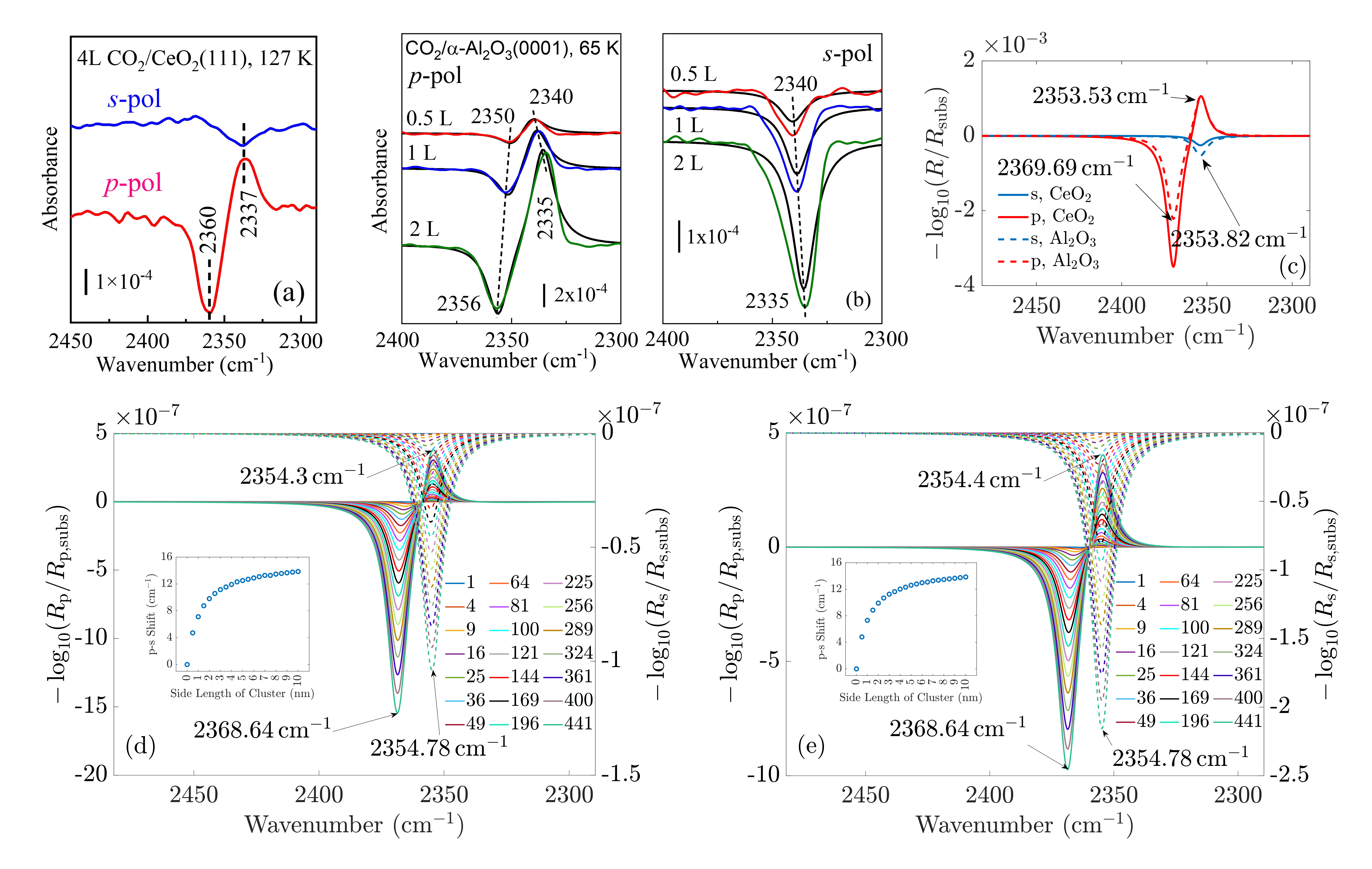}
	\caption{\textbf{(a)} Experimentally measured spectra of CO$_2$ adsorbed on CeO$_2$(111) and \textbf{(b)} $\alpha$-Al$_2$O$_3$(0001) single-crystal surfaces (the red, blue, and green curves represent fits based on the four-layer-model). \textbf{(c)} Simulated spectra of the corresponding situations. Experimentally and phenomenologically, it is shown in \textbf{(b)} that the exact position of the peaks and their difference depend on the amount of molecules used to cover the substrate. This can explain some disagreements between the simulated and the measured results. \textbf{(d)} Islands with finite size on CeO$_2$ and \textbf{(e)} on Al$_2$O$_3$. The shift between the s- and p-resonances is zero for one molecule. For increasing island sizes, the polarization-dependent shift increases.}
    \label{fig:CompExpToTheoryCO2CeriaAl2O3}
	\end{figure*}
\subsection{Islands of CO$_2$ molecules supported on a substrate}

In the following, we concentrate on CO$_2$ molecules. In Appendix~A.1, the experimental measurement is explained in more detail. In Figure~\ref{fig:CompExpToTheoryCO2CeriaAl2O3}, we compare the spectra for CO$_2$ molecules adsorbed on CeO$_2$(111) and  $\alpha$-Al$_2$O$_3$(0001) single-crystal surfaces, respectively. In general, the results for ceria and Al$_2$O$_3$ are very similar.
In Figure~\ref{fig:CompExpToTheoryCO2CeriaAl2O3}\textbf{(a)} and \textbf{(b)}, the experimentally measured spectra are shown. For comparison, Figure~\ref{fig:CompExpToTheoryCO2CeriaAl2O3}\textbf{(c)} displays the simulation results under the assumption of an infinitely extended lattice. The agreement between the positions of the absorbance peaks in the measured and simulated spectra is good. For the theoretically simulated spectra, the polarization-dependent shift amounts to $15.87\,\mathrm{cm}^{-1}$, which is considerably larger than that for CO molecules ($3\,\mathrm{cm}^{-1}$). This is expected, also the three/four-layer-model in combination with dispersion theory predicts that for larger coupling strengths the shift increases (see Figure~\ref{fig:CompExpToTheoryCO2CeriaAl2O3}\textbf{(b)}). \cite{CHABAL1988211} While the position and the shifts of the peaks slightly differ between the experimental and theoretical results, we observe the same trend in the measured and simulated spectra for both CeO$_2$ and Al$_2$O$_3$ for which the s-peak is very close to the p-peak at lower wavenumbers. The experiment and the phenomenological study for Al$_2$O$_3$ shown in Figure~\ref{fig:CompExpToTheoryCO2CeriaAl2O3}\textbf{(b)} reveal that the change of the amount of used molecules to cover the substrate shifts the peaks, explaining slight disagreements between experiment and theory. The phenomenological study is explained in more detail in Appendix~A.3.

At last, we study the optical response of the CO$_2$ molecular islands on the dielectric substrates as a function of their size. In Figures~\ref{fig:CompExpToTheoryCO2CeriaAl2O3}\textbf{(d)} and \textbf{(e)}, we observe, similar to the case of CO discussed above, that with increasing island size, the shift between the absorbance peaks in the s- and p-spectra increases until it is relatively close to the shift of the continuous adlayer. Again, we do not observe a polarization-dependent shift for a single adsorbate molecule. The study of the change of the coverage in Figure~\ref{fig:CompExpToTheoryCO2CeriaAl2O3}\textbf{(b)} demonstrates the same behavior as with the increasing number of used molecules in the experiment, which corresponds to smaller distances between the molecules and, therefore, larger interaction, the shift of the peaks increases. Together with the other results, this finding strongly supports the predictive power of our multi-scale modeling approach.

\section{Conclusion}
We considered the optical response of differently sized islands of CO and CO$_2$ molecules supported on two prototypic oxidic substrates, CeO$_2$(111) and $\alpha$-Al$_2$O$_3$(0001). Our \textit{ab initio} approach reproduces the experimentally observed shift between the absorbance peaks in the reflectance of s- and p-polarized plane waves. The experimentally measured and theoretically simulated results agree very well for an infinite adlayer. It was established that the shift value depends strongly on the properties of the adsorbates, but a dependence on the substrate is rather small. In the analysis of the finite systems, it was observed that the shift between the s- and the p-spectra vanishes for a single, isolated adsorbate molecule. However, already for islands consisting of four molecules, a shift is observed, which increases with increasing island size and converges to the value of the infinite systems already for island sizes of 10\,nm × 10\,nm. These results imply that with regard to the interpretation of infrared data recorded for oxide powders, size effects can be generally excluded for particle sizes above 10\,nm. 

\section*{Data availability statement}
The  data  that  support  the  findings  of  this  study  are  available  from  the  corresponding author upon reasonable request.
\section*{Conflicts of interest}
There are no conflicts to declare.

\section*{Acknowledgement}

M.K., D.B., C.W., and C.R. acknowledge support by the Deutsche Forschungsgemeinschaft (DFG, German Research Foundation) under Germany’s Excellence Strategy via the Excellence Cluster 3D Matter Made to Order (EXC-2082/1-390761711) and from the Carl Zeiss Foundation via the CZF-Focus@HEiKA Program. S.C., Z.Y., A.N., Y.W., and C.W. acknowledge support by the Deutsche Forschungsgemeinschaft (DFG, German Research Foundation) – Project-ID 426888090 – SFB 1441. S.C. is grateful for a Postdoc fellowship donated by the Helmholtz Association and China Postdoctoral Council (OCPC). M.K., C.H., and C.R. acknowledge funding by the Volkswagen Foundation. A.N., C.W. and C.R. acknowledge support by the Helmholtz Association via the
Helmholtz program “Materials Systems Engineering” (MSE). B.Z., M.N., and C.R. acknowledge support by the KIT through the “Virtual Materials Design” (VIRTMAT) project. We thank Dr. Ivan Fernandez-Corbaton for fruitful discussions about scattering simulation methods. M.K. and C.R. acknowledge support by the state of Baden-Württemberg through bwHPC and the German Research Foundation (DFG) through grant no. INST 40/575-1 FUGG (JUSTUS 2 Cluster) and the HoreKa supercomputer funded by the Ministry of Science, Research and the Arts Baden-Württemberg and by the Federal Ministry of Education and Research.

\appendix
\section{Experimental section}\label{Sec:Experiment}
The polarization-resolved IRRAS measurements were conducted in a dedicated ultrahigh vacuum (UHV) apparatus combining a state-of-the-art FTIR spectrometer (Vertex 80v, Bruker Optics GmbH, Germany) with a multi-chamber UHV system (PREVAC, Poland). \cite{Wang2017} Prior to the IRRAS experiments, the CeO$_2$(111) (SurfaceNet, Germany) and $\alpha$-Al$_2$O$_3$(0001) (Crystal GmbH, Germany) single crystals were prepared using standard procedures.  The ceria surface was pretreated by repeated cycles of sputtering with 1\,keV Ar$^+$ and annealing at 800\,K in an O$_2$ atmosphere of $1\times 10^{-5}$\,mbar to form a stoichiometric surface. The $\alpha$-Al$_2$O$_3$(0001) single crystal was pretreated by repeated cycles of sputtering with 2\,keV Ar$^+$ and annealing at 850\,K in an O$_2$ atmosphere of $1\times 10^{-5}$\,mbar. The cleanliness and oxidation states of the samples were monitored by grazing incidence XPS equipped with a R4000 electron energy analyzer (VG Scienta, Sweden). Exposure to CO and CO$_2$ at low temperatures was achieved by backfilling the IR chamber using a leak-valve-based directional doser connected to a tube (2\,mm in diameter) that terminated 3\,cm from the sample surface. The exposure was carried out at $1.3\times 10^{-9}$\,mbar in a few steps with a total exposure time of 40\,s. Since the hot-cathode ionization gauge, which was used for exposure control, is placed closed to the pump and in 50\,cm distance from the sample, the real exposure was, according to our internal calibtation, about 1000 times higher resulting in a coverage of 4\,Langmuir (L). The base pressure during acquisition of IR spectra was $~3\times 10^{-11}$\,mbar. Polarization-resolved IRRA spectra were recorded with both s- and p-polarized light at a fixed grazing incidence of 80$^{\circ}$. All spectra shown are difference spectra obtained by subtracting a reference spectrum recorded immediately before adsorption of CO$_2$ or CO at 127\,K (cooling with liquid nitrogen) or 65\,K (cooling with liquid helium), respectively.  
\section{T-matrix formalism}\label{Sec:TmatrixFormalism}
We use the T-matrix formalism to describe the scattering of light at the molecular islands. \cite{Waterman1965} The T-matrix relates the multipolar coefficients of the scattered electric field of an object to the multipolar coefficients of the incident field. In the frequency domain, one expands the total electric field outside a scattering object at point $\bm{r}$ as
\begin{align}\label{eq:EExpan}
\begin{split}
    \bm{E}(\bm{r})=\sum_{l=1}^{\infty}\sum_{m=-l}^{l}
    \left(
        a_{lm,\mathrm{N}}\bm{N}^{(1)}_{lm}(k_{\mathrm{h}}\bm{r})
        + a_{lm,\mathrm{M}}\bm{M}^{(1)}_{lm}(k_{\mathrm{h}}\bm{r})
    \right.
        \\
    \left.
        + c_{lm,\mathrm{N}}\bm{N}^{(3)}_{lm}(k_{\mathrm{h}}\bm{r})
        + c_{lm,\mathrm{M}}\bm{M}^{(3)}_{lm}(k_{\mathrm{h}}\bm{r})
    \right)\,.
    \end{split}
\end{align}
$\bm{N}^{(1)}_{lm}(k_{\mathrm{h}}\bm{r})$ and $\bm{M}^{(1)}_{lm}(k_{\mathrm{h}}\bm{r})$ are incident and $\bm{N}^{(3)}_{lm}(k_{\mathrm{h}}\bm{r})$ and $\bm{M}^{(3)}_{lm}(k_{\mathrm{h}}\bm{r})$ are scattered vector spherical waves. For regular waves, $\bm{N}$ and $\bm{M}$ are transverse magnetic (TM) and transverse electric (TE) modes, respectively. \cite{Fernandez-Corbaton:2020,Beutel:21} $k_{\mathrm{h}}=\omega\sqrt{\varepsilon_{\mathrm{h}}\mu_{\mathrm{h}}}$ is the wave number with the permeability $\mu_{\mathrm{h}}$ and the permittivity $\varepsilon_{\mathrm{h}}$ of the host medium. All quantities are frequency-dependent. However, we omit the frequency as an explicit argument for better readability. The expansion coefficients $a_{lm}$ and $c_{lm}$ are related by the T-matrix, 
\begin{align}\label{eq:DefTMat}
    \bm{c}
    =\mathbf{T}
    \bm{a}\,.
\end{align} 
To compute the dipolar T-matrix of a molecule, Equation~(6) from Ref.~ \cite{Fernandez-Corbaton:2020} is used,
\begin{align}\label{eq:PolToTMat}
\begin{split}
\begin{pmatrix}
\mathbf{T}_{\mathrm{NN}}&\mathbf{T}_{\mathrm{NM}}\\
\mathbf{T}_{\mathrm{MN}}&\mathbf{T}_{\mathrm{MM}}
\end{pmatrix}&
=\frac{\mathrm{i}c_{\mathrm{h}}Z_{\mathrm{h}}k_{\mathrm{h}}^3}{6\pi}\times\\
&\times\begin{pmatrix}
\mathbf{C}\left(\bm{\alpha}_{\mathrm{ee}}\right)\mathbf{C}^{-1}&\mathbf{C}\left(-\mathrm{i}\bm{\alpha}_{\mathrm{em}}/Z_{\mathrm{h}}\right)\mathbf{C}^{-1}\\
\mathbf{C}\left(\mathrm{i}\bm{\alpha}_{\mathrm{me}}/c_{\mathrm{h}}\right)\mathbf{C}^{-1}&\mathbf{C}\left(\bm{\alpha}_{\mathrm{mm}}/(c_{\mathrm{h}}Z_{\mathrm{h}})\right)\mathbf{C}^{-1}
\end{pmatrix}\,,
\end{split}
\end{align}
which relates the dipolar polarizability tensors $\bm{\alpha}_{\nu \nu'}$ to the T-matrix with simple matrix multiplications. $\mathbf{C}$ is a unitary matrix to transform the polarizability tensors from the Cartesian to the spherical basis. $\bm{\alpha}_{\nu \nu'}$ are complex $3\times 3$ dipolar polarizability tensors, which are computed with TDDFT. \cite{Fernandez-Corbaton:2020,SURMOFCavity,https://doi.org/10.1002/adfm.202301093,Kehry.Franzke.ea:Quasirelativistic.2020} $c_{\mathrm{h}}=1/\sqrt{\varepsilon_{\mathrm{h}}\mu_{\mathrm{h}}}$ is the speed of light in the medium surrounding the molecules and $Z_{\mathrm{h}}=\sqrt{\mu_{\mathrm{h}}/\varepsilon_{\mathrm{h}}}$ is the corresponding wave impedance. 

The optical response of a two-dimensional lattice of molecules is obtained by solving a multi-scattering problem, where the expansion coefficients $\bm{a}_0$ of the primary incident field and the expansion coefficients $\bm{c}_{0,\mathrm{tot}}$ of the scattered field of a molecule at the origin of the lattice are determined \cite{Beutel:21}:
\begin{align}\label{eq:MultiScat}
    \bm{c}_{0,\mathrm{tot}}=\left(\mathds{1}-\mathbf{T}\sum_{\bm{R}\neq 0}\mathbf{C}^{(3)}(-\bm{R})\mathrm{e}^{\mathrm{i}\bm{k}_{\parallel}\bm{R}}\right)^{-1}\mathbf{T}\bm{a}_0\,.
\end{align}
In Equation~(\ref{eq:MultiScat}), $\bm{R}$ is a two-dimensional lattice vector, $\bm{k}_{\parallel}$ is the component of the illuminating plane wave parallel to the lattice, and $\mathbf{C}^{(3)}(-\bm{R})$ is a matrix which translates scattered vector spherical waves to incident vector spherical waves.

Additionally to solving Equation~(\ref{eq:MultiScat}), one computes Q-matrices defined in Equations~(6)-(9), (12a,b) in Ref.~ \cite{Beutel:21} to describe the substrate and the interfaces between the different layers. The combination of the different Q-matrices enables the computation of the reflectance of the entire device.

To describe the scattering of an adsorbate island consisting of $N$ molecules, the molecular island is represented by T-matrices $\mathbf{T}^{nj}$, which relate the expansion coefficients of the incident field $\bm{a}^{j}$ at particle $j$ to the expansion coefficients of the total scattered field $\bm{c}^{n}$ of particle $n$, 
\begin{align}
\bm{c}^{n} = \sum_{j=1}^N \mathbf{T}^{nj} \bm{a}^{j}\,,
\end{align}
see Equation~(76) from Ref.~ \cite{MISHCHENKO1996}, Equation~(16) from Ref.~ \cite{Mackowski:94}, and Equation~(19) from Ref.~ \cite{BEUTEL2024109076}. In treams, \cite{Beutel:21,treams,BEUTEL2024109076} the T-matrices $\mathbf{T}^{nj}$ summarized into a single local T-matrix can also be used to compute the scattering response of a two-dimensional superlattice of islands with Equation~(\ref{eq:MultiScat}).

\section{Three/four-layer-model study of CO$_2$ on Al$_2$O$_3$}\label{sec:3LayerModel}
For the simulations and the fits of the experimental spectra, we assumed either a three-layer model for CO absorbed on ceria (semi-infinite incidence medium/absorbed CO/semi-infinite substrate) or a four-layer-model for CO$_2$ adsorbed on Al$_2$O$_3$ (semi-infinite incidence medium/absorbed CO$_2$/substrate/semi-infinite exit medium). Incidence medium and exit medium were assumed to be vacuum. Ceria was modelled with a scalar dielectric function, whereas Al$_2$O$_3$ was assumed to be oriented with the c-axis normal to the surface. For the adsorbed layer we employed a matrix approach to compute an orientation-averaged dielectric tensor according to Equation~(\ref{Eq:EpsLayerModel}):
\begin{align}\label{Eq:EpsLayerModel}
\begin{split}
    \langle \bm{\varepsilon}_{\mathrm{r},j}\rangle &=\int_0^{2\pi}\mathbf{A}\cdot\begin{pmatrix}
        \varepsilon_{\mathrm{r},j,a}&0&0\\
        0&\varepsilon_{\mathrm{r},j,a}&0\\
        0&0&\varepsilon_{\mathrm{r},j,c}
    \end{pmatrix}\cdot \mathbf{A}^{-1}\mathrm{d}\varphi\, ,\\
    \mathbf{A}&=\begin{pmatrix}
        \cos(\varphi)&-\cos(\theta)\sin(\varphi)&\sin(\theta)\sin(\varphi)\\
        \sin(\varphi)&\cos(\theta)\cos(\varphi)&-\sin(\theta)\cos(\varphi)\\
        0&\sin(\theta)&\cos(\theta)
    \end{pmatrix}\, .
\end{split}
\end{align}
From Equation~(\ref{Eq:EpsLayerModel}) results an effective medium that is uniaxial and whose optical axis is oriented perpendicular to the interface. The principal dielectric functions, $\varepsilon_{\mathrm{r},j,a}$ and $\varepsilon_{\mathrm{r},j,c}$, were generated by the classical damped harmonic oscillator model, 
\begin{align}
    \varepsilon_{\mathrm{r},j}(\tilde{\nu})=\varepsilon_{\mathrm{r},j,\infty}+\sum_{i=1}^N\frac{S_{ji}^2}{\tilde{\nu}_{ji}^2-\tilde{\nu}^2-i\tilde{\nu}\gamma_{ji}}\, ,
\end{align}
where $S_{ji}^2$ is the oscillator strength of the $i$-th oscillator, $\tilde{\nu}_{ji}$ its position and  $\gamma_{ji}$ the damping (which was set to 5\,cm$^{-1}$ for the TDDFT calculations). $\varepsilon_{\mathrm{r},j,\infty}$ represents the dielectric background which is due to modes in the UV/Vis spectral region. 

For the calculation of the reflectances, a unified 4×4 matrix formalism was employed. \cite{Mayerhöfer_2006,Mayerhöfer2024} The experimental absorbance spectra $-\log_{10}(R_{\mathrm{s}}/R_{\mathrm{s,subs}})$ and $-\log_{10}(R_{\mathrm{p}}/R_{\mathrm{p,subs}})$ were baseline corrected by subtracting line functions that were fitted to the data points in non-absorbing regions between 2280 and 2420\,cm$^{-1}$. For the fits of the absorbance bands of CO$_2$ all 7 parameters shown in Table~\ref{table:ParamsLayerModel} were allowed to vary freely. The distance between maximum and minimum in p-polarized spectra is proportional to the oscillator strength. Correspondingly, the oscillator strengths increase with increasing number density of CO$_2$-molecules. The damping of the oscillator decreases, but not substantially. The redshift of the oscillator position could be a polarization effect as it increases with number density. \cite{Mayerhofer:22} It is, however, somewhat surprising that neither the $\varepsilon_{\mathrm{r},a,\infty}$ nor $\varepsilon_{\mathrm{r},c,\infty}$ change with increasing density, since also the oscillator strengths in the UV/Vis should correlate with changes of density of CO$_2$. The angles $\theta$ relative to the surface normal stay approximately constant, while the thickness of the layers $d$ increases. However, it is not clear for a monomolecular or bimolecular layer what this value actually refers to, since we used the same 4x4 matrix formalism that is usually applied for homogenous solid layers. The oscillator parameters of Al$_2$O$_3$ were taken over from Ref.~ \cite{PhysRev.132.1474}.

\begin{table}
\begin{center}
  \fontsize{7.6}{12}\selectfont
  \caption{Parameters based on the fits of the experimental spectra of CO$_2$ on Al$_2$O$_3$.}
  \label{table:ParamsLayerModel}
  \begin{tabular}{rccccccc}
    \hline
      &$S$/cm$^{-1}$&$\gamma$/cm$^{-1}$&$\nu_0$/cm$^{-1}$&$\varepsilon_{\mathrm{r},a,\infty}$&$\varepsilon_{\mathrm{r},c,\infty}$&$\theta$/deg&$d$/\textup{\AA} \\
    \hline
0.5\,L&519.3&14.28&2340.8&1.61&2.38&56.5&1.31\\
1\,L&645.5&13.09&2338.8&1.67&2.35&57.3&1.64\\
2\,L&845.5&12.72&2335.7&1.69&2.37&57.5&2.05\\
    \hline
  \end{tabular}
  \end{center}
\end{table}

\bibliography{bibliographyArxiv}

\end{document}



\title{Polarization-dependent effects in vibrational absorption spectra of 2D finite-size adsorbate islands on dielectric substrates}

\author{Benedikt Zerulla}
\affiliation{Institute of Nanotechnology,
Karlsruhe Institute of Technology (KIT),
Kaiserstr. 12, 76131 Karlsruhe, Germany}
\author{Marjan Krstić}
\affiliation{Institute of Theoretical Solid State Physics,
Karlsruhe Institute of Technology (KIT),
Kaiserstr. 12, 76131 Karlsruhe, Germany}
\author{Shuang Chen}
\affiliation{Institute of Functional Interfaces,
Karlsruhe Institute of Technology (KIT),
Kaiserstr. 12, 76131 Karlsruhe, Germany}
\author{Zairan Yu}
\affiliation{Institute of Functional Interfaces,
Karlsruhe Institute of Technology (KIT),
Kaiserstr. 12, 76131 Karlsruhe, Germany}
\author{Dominik Beutel}
\affiliation{Institute of Theoretical Solid State Physics,
Karlsruhe Institute of Technology (KIT),
Kaiserstr. 12, 76131 Karlsruhe, Germany}
\author{Christof Holzer}
\affiliation{Institute of Theoretical Solid State Physics,
Karlsruhe Institute of Technology (KIT),
Kaiserstr. 12, 76131 Karlsruhe, Germany}
\author{Markus Nyman}
\affiliation{Institute of Nanotechnology,
Karlsruhe Institute of Technology (KIT),
Kaiserstr. 12, 76131 Karlsruhe, Germany}
\author{Alexei Nefedov}
\affiliation{Institute of Functional Interfaces,
Karlsruhe Institute of Technology (KIT),
Kaiserstr. 12, 76131 Karlsruhe, Germany}
\author{Yuemin Wang}
\affiliation{Institute of Functional Interfaces,
Karlsruhe Institute of Technology (KIT),
Kaiserstr. 12, 76131 Karlsruhe, Germany}
\author{Thomas G. Mayerhöfer}
\affiliation{Leibniz Institute of Photonic Technology,
Albert-Einstein-Str. 9, 07745 Jena, Germany}
\author{Christof Wöll}
\affiliation{Institute of Functional Interfaces,
Karlsruhe Institute of Technology (KIT),
Kaiserstr. 12, 76131 Karlsruhe, Germany}
\author{Carsten Rockstuhl}
\affiliation{Institute of Nanotechnology,
Karlsruhe Institute of Technology (KIT),
Kaiserstr. 12, 76131 Karlsruhe, Germany}
\affiliation{Institute of Theoretical Solid State Physics,
Karlsruhe Institute of Technology (KIT),
Kaiserstr. 12, 76131 Karlsruhe, Germany}

\date{\today}

\begin{abstract}
\end{abstract}

\pacs{}

\maketitle 


\begin{figure}[H]
\centering
     
	\includegraphics[width=0.85\textwidth]{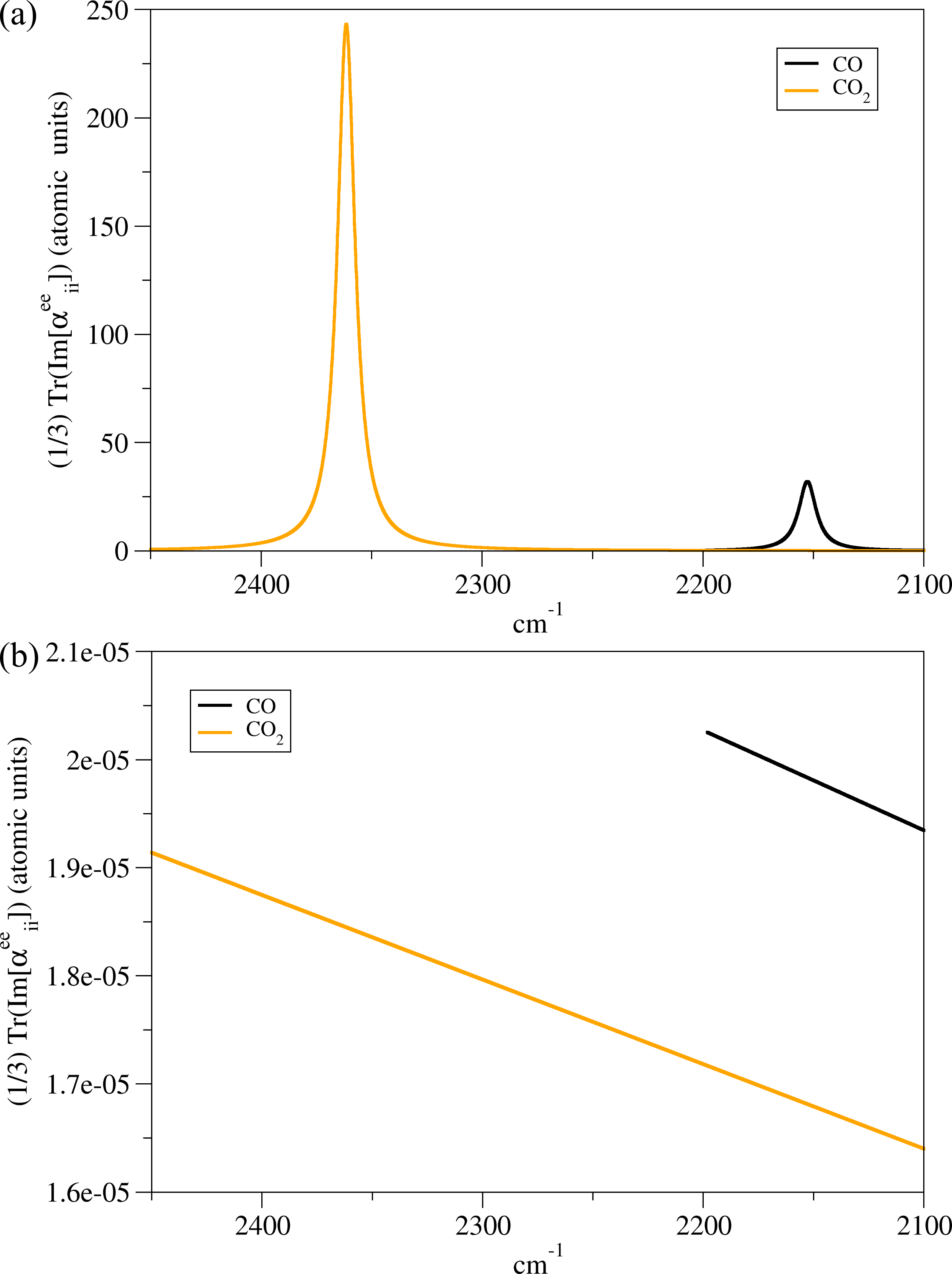}
\hspace{0cm}
	\caption{\textbf{(a)} Visualized IR absorption of the CO and CO\textsubscript{2} molecule simulated from the trace of the calculated vibrational dynamic polarizabilities using DFT method. \textbf{(b)} Electronic absorption of CO and CO\textsubscript{2} molecule in the infra-red part of the spectrum, representing long-range tail of the damped electronic transition in the UV part of the spectrum.}
    \label{fig:DFT_pol}
	\end{figure}
\newpage

\bibliography{bibliographyArxiv}